\documentclass[12pts,aps,twocolumn,prd,showpacs,nofootinbib,preprintnumbers]{revtex4-1}
\usepackage{bm}
\usepackage{latexsym}
\usepackage{natbib}
\usepackage{url}
\usepackage{dcolumn}
\usepackage{color}
\usepackage{amsfonts,amssymb,amsmath}
\usepackage{graphicx,epsfig}
\usepackage{psfrag}
\usepackage{subfigure}
\usepackage{natbib}
\usepackage{array,makecell}

\begin{document}

\title{Light deflection angle through velocity profile of galaxies in $f(R)$ model}

\author{ Vipin Kumar Sharma}
\email[]{\textcolor[rgb]{1.00,0.00,1.00}{vipinastrophysics@gmail.com}}
\author{ Bal Krishna Yadav}
\email[]{\textcolor[rgb]{1.00,0.00,1.00}{balkrishnalko@gmail.com}}
\author{Murli Manohar Verma}
\email[]{\textcolor[rgb]{1.00,0.00,1.00}{sunilmmv@yahoo.com}, \textcolor[rgb]{1.00,0.00,1.00}{murli.manohar.verma@cern.ch}}

\affiliation{$^{*, \dagger,\ddagger}$Department of Physics, University of Lucknow, Lucknow 226 007, India \\ $^{\ddagger}$Theoretical Physics Division, CERN, CH-1211 Geneva 23, Switzerland}
\date{\today}
\begin{abstract}
We explore a  new realisation of  the galactic scale dynamics via gravitational lensing phenomenon in power-law $f(R)$ gravity theory of the type $f(R)\propto R^{1+\delta}$ with $\delta<<1$ for interpreting the clustered dark matter effects. We utilize the single effective point like potential (Newtonian potential + $f(R)$ background potential) obtained under the weak field limit to study the combined observations of galaxy rotation curve beyond the optical disk size and their lensing  profile in $f(R)$ frame work.  We calculate the magnitude of light deflection angle  with  the characteristic length scale (because of Noether symmetry in $f(R)$ theories) appearing  in the effective $f(R)$ rotational velocity profile of  a  typical galaxy with the model parameter $\delta \approx O(10^{-6})$ constrained in previous work. For instance, we work with the two nearby controversial galaxies  NGC 5533 and NGC 4138 and explore  their galactic features by  analysing the lensing angle profiles in $f(R)$ background.
We  also  contrast the magnitudes of $f(R)$  lensing  angle profiles and the relevant parameters of such galaxies  with the generalised pseudo-isothermal galaxy halo model and find consistency. 


 \end{abstract}
\maketitle{}

\section{\label{1}Introduction}
The conventional approach to  explain  the observed cosmological consequences at different redshifts  has been proposed  by assuming  the presence of some hidden sectors called dark matter ($\approx$ 27 {\%}) and the dark energy ($\approx$ 68 {\%})\textcolor[rgb]{0.00,0.00,1.00}{\cite{b1}} in the  Einstein's General Relativity (GR). It is found  that at the scale of solar system, GR produces  the precise astrophysical results without requiring any clustered and pressureless non-standard matter \textcolor[rgb]{0.00,0.00,1.00}{\cite{b2,b3,b4}}.  Also, the gravitational deflection angle of light due to  Sun provided   us  with   the first unique observational test of GR.  However,  beyond the scale of Solar System, especially in  the  large galactic regions,    it does not succeed much without such dark sector. \textcolor[rgb]{0.00,0.00,1.00}{\cite{b5,b6,b7}}.   Among the observations  at  galactic scales, the rotational velocity profile (inner and outer rotation curves) of a typical galaxy after H${\alpha}$ and HI observations \textcolor[rgb]{0.00,0.00,1.00}{\cite{b8,b9}} and the lensing angles  due to individual galaxies  and  cluster of galaxies collectively  \textcolor[rgb]{0.00,0.00,1.00}{\cite{b10,b11,b12}} are the  most profound ones.  Both  of these astrophysical observations predict the presence of pressureless clustered dark matter at low cosmological redshifts.  Actually, from the dynamical point of view, there are three  basic types of cosmological dark matter (DM) content:  (i) local DM in the disk of galaxies (ii) DM in the halos  or coronae of galaxies and system of galaxies, and   (iii) non-clustered DM.  There have  been significant theoretical efforts to explain the effects of such clustered DM beyond the standard Einstein's  gravity theory \textcolor[rgb]{0.00,0.00,1.00}{\cite{b13,b14,b15}} at the galactic scales.  Since  till date, the particulate form of dark matter is still unrevealed through various particle physics experimental investigations \textcolor[rgb]{0.00,0.00,1.00} {\cite{b16}}.
Thus, to deal with the unexplained hidden dark sector along with the normal standard matter fields,  a modification in the Einstein's spacetime gravity theory is preferred  and hence we have a different  set of field equations and different cosmological interpretation.

The most general and simplest modification in the standard Einstein's gravity theory is introduced   by replacing the Einstein-Hilbert gravity Lagrangian density by a functional form of the Ricci scalar curvature $(R)$ \textcolor[rgb]{0.00,0.00,1.00}{\cite{b15,b17,b18,b19,b20}}. Such a modification in general invokes some additional terms (extra scalar degree of freedom attached with the Ricci scalar curvature) other than the usual terms in the Einstein's   equations. These extra terms can explain the several cosmological problems subject  to the extent of modification.

It is to be noted that any modification in the Einstein-Hilbert gravity action integral will lead to a different order gravity theory. The order of any gravity theory is determined via the specific Lagrangian density. Because the d' Alembertian operator  introduces two more derivative terms into the modified Einstein's field equations and also if the function i.e., $f(R)$ is differentiable at least up to the second order, we have a fourth order gravity theory. One of the most important consequences   of the power-law $f(R)$ modification  is the emergence of new characteristic length scale due to Noether symmetry \textcolor[rgb]{0.00,0.00,1.00}{\cite{b21,b22,b23}}.  It offers the possibility to obtain the viable rotation curves without dark matter and the extent of modification required for this has been  also discussed by several researchers \textcolor[rgb]{0.00,0.00,1.00}{\cite{b24,b25,b26}}. In metric formalism,  the gravity theory of order 2$N$  possesses $N$ characteristic length scale \textcolor[rgb]{0.00,0.00,1.00}{\cite{b27}}. For instance, Einstein's GR is a second-order metric theory of gravity,  and so  it has one characteristic length scale  i.e., Schwarzschild length scale. $f(R)$ formulation  is built on a fourth-order theory and it must have two characteristic length scales,  of which one is the Schwarzschild length scale and the other  is  the  $f(R)$ characteristic length scale appearing   generally because of the extent of power-law modification in the  curvature term.

There are several ways to study the vast field of modified Einstein gravity theory in an alternate manner. Some of them are: MOdified Newtonian Dynamics (MOND), Brans-Dicke (B-D) theory, Klauza-Klein (K-K) theory, Braneworld models, $f(R)$ theory,  $f(T)$ theory, $f(R,T)$ theory, Scalar Tensor Vector Gravity (STVG) theory,  mimetric gravity theory etc. Such theories have a potential to explain the cosmological dark sector problems successfully instead of  following the traditional approach.  Also, such theories have  gained importance due to their general scope   in explaining the physics of the universe from the very early epochs   up to  the   present accelerated phase without facing serious cosmological problems  viz.,    fine tuning problem and coincidence problem as faced by standard Lambda Cold Dark Matter ($\Lambda$CDM) model \textcolor[rgb]{0.00,0.00,1.00}{\cite{b28}}. For instance,  one  can  handle    the  fine tuning problem by working with vacuum $f(R)$ field equations.  More recently, $f(R)$ gravity has  drawn attention in discussion of  galaxy formation \textcolor[rgb]{0.00,0.00,1.00}{\cite{b29}} and  the explanation of dark matter like effects at different redshifts \textcolor[rgb]{0.00,0.00,1.00}{\cite{b30}}. The particle picture  of scalaron  as a dark matter candidate has also  been  explored  \textcolor[rgb]{0.00,0.00,1.00}{\cite{b031}}.

In the present paper,  we consider the modified gravity Lagrangian density of the type, $f(R)\propto R^{1+\delta}$;  where  $\delta>0$. Due to the inheritance of the scalar degrees of freedom in the $f(R)$ gravity theory, the galactic rotation velocity profile as well as the lensing angle profile of a typical galaxy must get altered from those in GR. Such alteration can explain the observed galactic rotational velocity profiles without   demanding  the existence   of dark matter in   contrast to GR theory \textcolor[rgb]{0.00,0.00,1.00}{\cite{b24,b25,b26}}. The extent of deviation $\delta$ for such $f(R)$ model can be constrained at the galactic scales via tracing of typical spiral galaxy outer rotation curves after HI observations  to be $O(10^{-6})$ \textcolor[rgb]{0.00,0.00,1.00}{\cite{b31}} which is consistent with the results explored in \textcolor[rgb]{0.00,0.00,1.00}{\cite{b310}}.

Therefore,  we investigate  the   influence  of  the scalaron  background effect  on the profile of lensing angle.  We may express  the  deflection angle  under the weak field limit in $f(R)$  gravity theory as $\hat{\alpha}=\frac{4GM}{c^2\xi}(1\pm\gamma)$, where $\xi$ is the two dimensional impact parameter and $\gamma$ is  the contribution due to the modification of the spacetime \textcolor[rgb]{0.00,0.00,1.00}{\cite{b32,b33}}. Further,  as   the   power-law   potential has been  investigated for the galactic rotation velocity and lensing phenomena \textcolor[rgb]{0.00,0.00,1.00}{\cite{b24}}, \textcolor[rgb]{0.00,0.00,1.00}{\cite{b32}},    we argue that their calculations  in $f(R)$  background obtained for  the same potential and value of $\delta$  \textcolor[rgb]{0.00,0.00,1.00}{\cite{b31}} would  yield some significant  information about the galactic dynamics. Although the rotation velocity profiles of different galaxies are proved to be the successful tool for the determination of  mass distribution,  we focus on the observations of rotation curve behaviour beyond the optical disk size for a  few galaxies in the modified gravity   background and calculate the magnitude of deflection   angles  by  using   the $f(R)$ characteristic length scale  obtained through the study of  outer rotation curves.    We  model the extent  of galactic  dark matter halo as  a  scalaron cloud.  Thus,  we  regard such $f(R)$ characteristic length in  determining   the   halo size of scalaron cloud in the power-law $f(R)$ gravity theory  instead of dark matter  halo.

 We discuss the galactic dynamics for nearby galaxy in  $f(R)$    cosmological background via calculating the formula of light deflection angle according to \textcolor[rgb]{0.00,0.00,1.00}{\cite{b32}}. In \textcolor[rgb]{0.00,0.00,1.00}{\cite{b34}}, the analysis of lensing angle profile according to the generalised pseudo-isothermal dark matter galaxy halo model is done for different nearby galaxies without any interpretation about the extent of halo size.  We study the two controversial galaxies i.e., NGC 5533 and NGC 4138 via the observations of rotational velocity profile beyond the optical disk size and lensing angle profile in $f(R)$ background.    A comparison   of  their   lensing  angle  profile   with the generalised pseudo-isothermal dark matter   halo model and interpret their magnitude and also discuss some important implications.

In Section \textcolor[rgb]{1.00,0.00,0.00}{(II)}, we discuss the basics of $f(R)$ dynamical field equations in vacuum and write the effective potential due to a massive spherically symmetric source in $f(R)$ background following our previous work \textcolor[rgb]{0.00,0.00,1.00}{\cite{b31}}. In Section \textcolor[rgb]{1.00,0.00,0.00}{(III)}, We explore the galactic dynamics via the $f(R)$ rotational velocity profiles for the nearby galaxies (under controversy due to their declining rotation curve) NGC 5533 and NGC 4138 and obtain their respective galactic length scale free parameter $r_0$ with $\delta\approx O(10^{-6})$. The formula of net light deflection angle in $f(R)$ background is obtained for the effective point like potential in Section \textcolor[rgb]{1.00,0.00,0.00}{(IV)} and explore further the lensing angle profiles for the said galaxies with their observed outer rotation curve profiles to obtain their lensing angle magnitude in $f(R)$ background. In Section \textcolor[rgb]{1.00,0.00,0.00}{(V)}, the magnitude of the light deflection angle for the same galaxies is explored with the generalised pseudo-isothermal dark matter galaxy model which is compared with the results of the net light deflection angle in $f(R)$ background and also its important implications are discussed. We conclude and discuss the results in Section \textcolor[rgb]{1.00,0.00,0.00}{(VI)}.  Throughout the paper, we follow the signature of the spacetime metric as ($-$,+,+,+) and indices $\mu$ (or $\nu$) =(0,1,2,3).

\section{\label{2}  $f(R)$ Dynamical equations and effective potential}
In order to study the modified effect of gravity under the Friedmann-Lemaitre-Robertson-Walker (FLRW) spatially flat background metric,
\begin{eqnarray}
ds^{2}= -dt^{2} + a^2(t)[dr^2 + r^2 (d\theta^{2} + \sin^{2}\theta d\phi^{2})] \label{a1},\end{eqnarray}
where $a(t)$ is the cosmological time dependent scale factor and ($r$,$\theta$,$\phi$ ) are the usual spherical coordinates, it is necessary that the function of Ricci scalar curvature in the Einstein-Hilbert action integral is dynamical i.e.,  $f(R) \neq R$. Here, we work with $f(R)=\frac{R^{1+\delta}}{R_{c}^{\delta}}$ type model with $\delta<<1$ at least at the galactic scales for the explanation of dark matter problem.\\
The four dimensional spacetime Einstein-Hilbert action of gravity in the modified theory is written in the absence of matter Lagrangian density in units of c=$\hbar$=1 as,
 \begin{eqnarray}
\mathcal{A}= \frac{1}{2}\int d^{4}x \sqrt{-g} \left[\frac{1}{8\pi G_N}f(R)\right] \label{a2},\end{eqnarray}
where $g$ is the determinant of the metric tensor $g_{\mu\nu}$ and $8\pi G_N$ is the Einstein's gravitational constant with  Newtonian gravitational constant $G_N$.\\
Now, to study the dynamics of such $f(R)$ gravity theory, we vary the action \textcolor[rgb]{0.00,0.00,1.00}{(\ref{a2})} w.r.t the metric tensor, which gives the dynamical field equations,
\begin{eqnarray}
F(R) R_{\mu\nu} -\frac{f(R)g_{\mu\nu}}{2} -{\nabla_\mu} {\nabla_\nu} F(R)+\nonumber\\ g_{\mu\nu}\Box F(R)=0 \label{a3},\end{eqnarray}
where  $F(R)$ is the first derivative of $f(R)$ w.r.t $R$, ${\nabla_\mu}$ is the covariant derivative associated with the Levi-Civita connection of the metric and $\Box\equiv {\nabla_\mu} {\nabla^{\mu}}$.

We can express the modified field equations in a formal way i.e., in tensor form and in the absence of energy-momentum tensor of standard matter as
\begin{eqnarray}
G_{\mu\nu}=\frac{8\pi G_N}{F(R)} [T^{(c)}_{\mu\nu}] \label{a4},\end{eqnarray}
where
\begin{eqnarray}
T^{(c)}_{\mu\nu}=\frac{1}{8\pi G_N}[\frac{1}{2} g_{\mu\nu}f(R)-\frac{R}{2} g_{\mu\nu} F(R)+\nonumber\\ \nabla_\mu\nabla_\nu F(R)- g_{\mu\nu} \Box F(R)] \label{a5},\end{eqnarray}
is the energy-momentum tensor of the spacetime curvature and $G_{\mu\nu}(= R_{\mu\nu}-\frac{R}{2} g_{\mu\nu})$, is the Einstein tensor.

In our previous work \textcolor[rgb]{0.00,0.00,1.00}{\cite{b31}}, we have obtained the modified effective potential in the weak field limit generated by the point-like source of mass $M$ in the $f(R)$ background and explored  the flat rotational velocity curves beyond the optical disk in the galaxy dark matter halo region.  Solving the vacuum field equations in the weak field limit for the discussion of nearby galaxies with the generalized spacetime metric (Schwarzschild-like) gives the modified effective potential for our model as,
\begin{eqnarray}
{V_{eff}}\approx-\frac{G_N M}{r}-\frac{{H_0}^2}{2 r^2}\left[ \chi r^{\frac{2+4\delta}{1+\delta}}-(\chi-1){r_0}^{\frac{2+4\delta}{1+\delta}}\right] ^{\frac{2(1+\delta)}{1+2\delta}} \label{a6}.\end{eqnarray}
The appearance of the first term in above equation \textcolor[rgb]{0.00,0.00,1.00}{(\ref{a6})} is due to the Newtonian potential of the gravitating source and the second term is due to the contribution of dynamical $f(R)$ cosmological background with ${r_0}$ as the $f(R)$ characteristic length scale parameter at the galactic scales according to \textcolor[rgb]{0.00,0.00,1.00}{\cite{b21,b22,b23,b27}}, which is a fundamental feature of the power law $f(R)$ gravity and constant  $\chi=\frac{R_0}{12 {H_0}^2}$. The Newtonian profile can be approximated  as we go back to GR since the relevant parameters vanishes in the source free limit.\\

In \textcolor[rgb]{0.00,0.00,1.00}{\cite{b24,b32}}, the unique effective potential of modified gravity model can be used for the explanation of the flatness of rotation curves as well as for the study of the light deflection angle. Here, we attempt to explore the light deflection angle from the flatness rotation velocity profiles of typical galaxies.

\section{\label{3} Rotational velocity profiles in $f(R)$ theory }

We trace the test mass beyond the visible disk region for the nearby galaxy via the effective $f(R)$ rotational velocity profile in \textcolor[rgb]{0.00,0.00,1.00}{\cite{b31}}. It is given as
\begin{eqnarray}
{v}^2\simeq{\frac {G_N M}{r}+\left( \frac{1}{2} \right) ^{(\frac{2+2\delta}{1+2\delta})}{H_0}^2\ r^2\ \left[ 1-\left( \frac{r_0}{r}\right)^{2(\frac{1+2\delta}{1+\delta})}\right]} \times \nonumber\\ {\left[ 1+\left( \frac{r_0}{r}\right)^{2(\frac{1+2\delta}{1+\delta})}\right] ^{\frac{1}{1+2\delta}}}.\label{a7}
\end{eqnarray}
Here, in our investigation we include the new local value of $H_0$  \textcolor[rgb]{0.00,0.00,1.00}{\cite{b032}} with a 2.4${\%}$ determination, $H_0 = 73.02\pm1.79$ km s$^{-1}$ Mpc$^{-1}$, $ G_N = 4.3\times 10^{-6}$ kpc km$^{2}$ sec$^{-2}$ $M$ $_{\odot}^{-1}$, $\chi(=\frac{R_0}{12 H_0^2})\approx\frac{1}{2}$ and $r_0$ is the $f(R)$ galactic length scale free parameter for a gravitating system (galaxies) which offers the possibility to fit the rotation curves in $f(R)$ background \textcolor[rgb]{0.00,0.00,1.00}{\cite{b24,b25,b26}}.

We set the constraint on $\delta \approx O(10^{-6})$ for the explanation of dark matter like effects through the rotation curve profile of the test mass beyond the optical disk size of the galaxy\textcolor[rgb]{0.00,0.00,1.00}{\cite{b31}}.
The value of $r_0$  is   obtained for the typical nearby galaxies,  NGC 5533 and NGC 4138 (addressed as controversial galaxies in literatures because of their declining rotation curve profile after HI observations) via tracing their approximate flat rotation curve outside the optical disk size as because the flatness paradox is mainly associated with such region.

The sample (first four columns) is presented  in Table \textcolor[rgb]{0.00,0.00,1.00}{I}.\\
\begin{table}[!ht]
\caption{Specifications of galaxies:  Galaxies for which the $f(R)$ galactic length scale parameter, i.e., $r_0$ is obtained via plotting the viable outer rotation velocity profiles by using equation \textcolor[rgb]{0.00,0.00,1.00}{(\ref{a7})}. The observed data is culled from \textcolor[rgb]{0.00,0.00,1.00}{\cite{b35}}.} 
\centering 
\begin{tabular}{c c c c c } 
\hline\hline 
Galaxy & \makecell{Optical radius\\ at isophotal level\\(kpc)}& \makecell{$M_{dyn}$\\$(\times10^{10}M_{\odot})$} & \makecell{$v_{asymp}$\\(km/s)} &  \makecell{ $r_0$ \\ (kpc)}\\ [0.5ex] 
\hline \\
NGC 5533 & 22.5 & 87.7 & 230 & 10$^{2.650}$ \\ 
NGC 4138 & 06.6 & 08.4  & 150 & 10$^{2.250}$ \\ [1ex] 
\hline 
\end{tabular}
\label{table:I} 
\end{table}

\begin{figure}[!h]
\centering  \begin{center} \end{center}
\includegraphics[width=0.44 \textwidth,origin=c,angle=0]{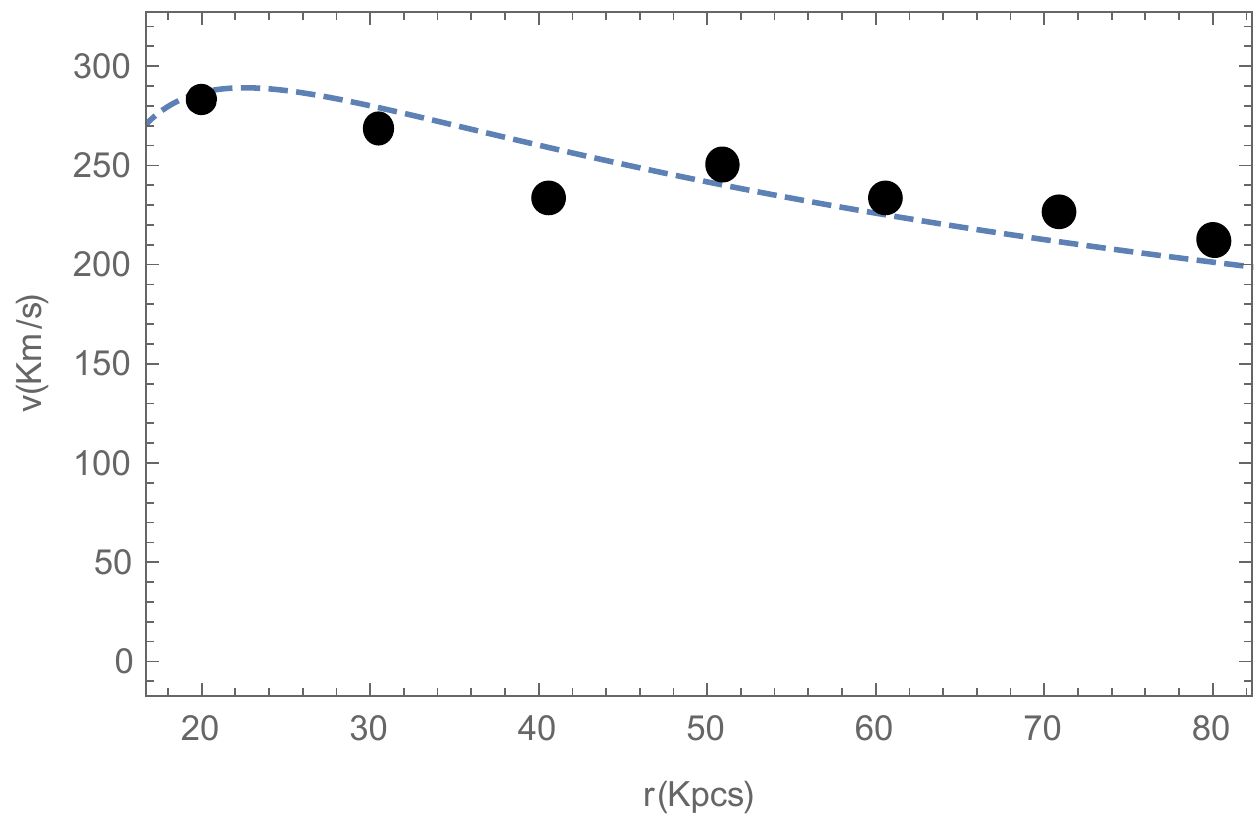}
\caption{\label{fig:p1} Theoretical galactic $f(R)$ rotation curve external to the typical visible end of galaxy NGC 5533. The black dots represent the observed data points due to HI \textcolor[rgb]{0.00,0.00,1.00}{\cite{b35}} . The curve shows the behaviour of test mass beyond the optical radius  for ${\delta\approx O(10^{-6}})$ with the galactic length scale free parameter $r_0=10^{2.650} \approx 446.6$ kpc.}\label{f1}
\end{figure}

\begin{figure}[!h]
\centering  \begin{center} \end{center}
\includegraphics[width=0.44 \textwidth,origin=c,angle=0]{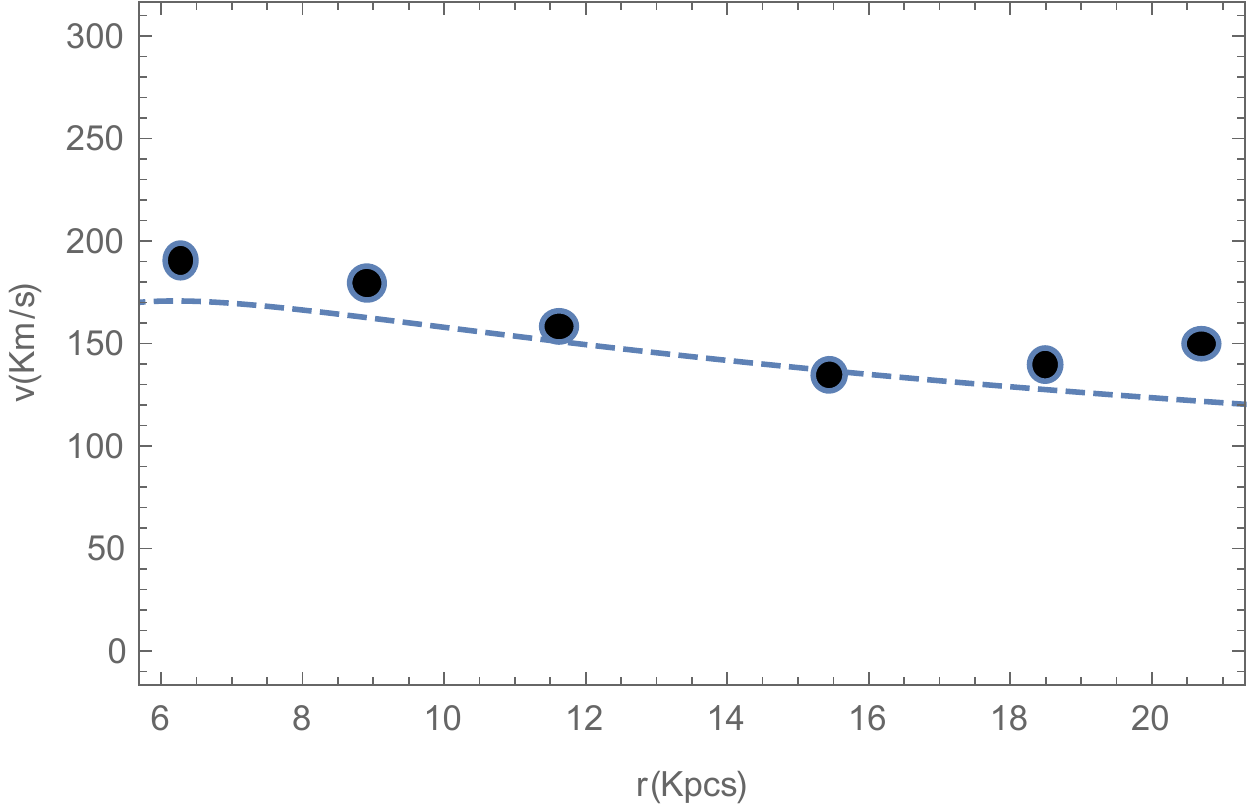}
\caption{\label{fig:p2}    Galactic rotation curve obtained for ${\delta\approx O(10^{-6}})$ with the $r_0=10^{2.250} \approx 177.828$ kpc external to the visible end of galaxy NGC 4138. The black dots represent the observed data points due to HI \textcolor[rgb]{0.00,0.00,1.00}{\cite{b35}}.}\label{f2}
\end{figure}

Fig. \textcolor[rgb]{0.00,0.00,1.00}{\ref{f1}} and  Fig. \textcolor[rgb]{0.00,0.00,1.00}{\ref{f2}} show the behaviour of the effective $f(R)$ rotational velocity curves beyond the optical size of galaxy. The fifth column $r_0$ in Table \textcolor[rgb]{0.00,0.00,1.00}{I} is obtained for them with $\delta \approx O(10^{-6})$ via tracing the observed rotational velocity plots   beyond the visible  boundaries  for dark matter like explanation.
From these  figures, it is clear that the effective $f(R)$ rotational velocity  becomes approximately constant beyond the outer visible region. Such profile is also observed through  HI observations. We  explore the implication of $f(R)$ characteristic length scale parameter $r_0$ for the galaxy halo. Therefore, we  study the lensing angle profiles for them in $f(R)$ background with the same respective $f(R)$ characteristic length scale as discussed next.

\section{\label{4} Magnitude of light deflection angle in $f(R)$ background}

Modifying the geometric theory of gravity  affects the underlined potential due to a spherically symmetric source of mass $M$ and hence will also affect the profile of lensing angle. According to \textcolor[rgb]{0.00,0.00,1.00}{\cite{b32,b36}}, the formal expression for the lensing angle in the modified gravity background  remains the  same as  in GR, but  with an effective potential.

The deflection in the path of a photon propagating along  $Z$ direction in $f(R)$ background having a spherical symmetric gravitating source of mass $M$ with an effective potential $V_{eff}$ given by equation \textcolor[rgb]{0.00,0.00,1.00}{(\ref{a6})} is  given as,
\begin{eqnarray}
\hat{\alpha}_{Net}=\frac{2}{c^2}\int_{-\infty}^{+\infty}{\nabla_{\bot}V_{eff}\ dz}\label{a8}.\end{eqnarray}
By  making use of $r^2\equiv x^2+y^2+z^2=\xi^2+z^2$ with $\nabla_{\bot}=\frac{\partial}{\partial\xi}$ and $\xi$ as a two dimensional impact parameter,  we obtain  through  some algebraic calculations
\begin{eqnarray}
\hat{\alpha}_{Net}={\frac{4G_NM}{c^2\xi}}-2\frac{H_0^2}{c^2}\sqrt{\pi}\xi^2 \left(\frac{1}{2}\right)^{\frac{2+2\delta}{1+2\delta}}\times \nonumber\\ \left[ \left(\frac{r_0}{\xi}\right)^4 \frac{\sqrt{\pi}}{2}-2\frac{\Gamma(-\frac{1}{2}+\frac{1}{1+\delta})}{\Gamma(\frac{1}{1+\delta})}\left(\frac{r_0}{\xi}\right)^{\frac{2}{1+\delta}}\right]  \label{a9}.\end{eqnarray}
 The first term of \textcolor[rgb]{0.00,0.00,1.00}{(\ref{a9})} is the standard GR,   while   the second term is the contribution due to $f(R)\neq R$.   Also, for $\delta=0$, we obtain the GR Lagrangian density  so  that its result can be recovered  because the extra scaling parameter must vanish for reasons mentioned in the Introduction.    Since  a   small  value of the $f(R)$ model parameter $\delta$ is preferred at the galactic scales for dark matter explanation, we    take $\delta \approx O(10^{-6})$ and use   \textcolor[rgb]{0.00,0.00,1.00}{(\ref{a9})} to obtain the magnitude of light deflection angle for the two galaxies.

The values of $f(R)$ galactic length scale free parameter ${r_0}$ for different galaxies is interpreted in Table \textcolor[rgb]{0.00,0.00,1.00}{I} via tracing the test mass beyond the outer visible region of galaxies by using equation \textcolor[rgb]{0.00,0.00,1.00}{(\ref{a7})}.  Further,  we  use it  in equation \textcolor[rgb]{0.00,0.00,1.00}{(\ref{a9})} for discussing the plots of the net light deflection angle w.r.t the scaled impact parameter i.e., $\frac{\xi}{r_0}$ in $f(R)$ background and to obtain its magnitude.
\begin{figure}[!h]
\centering  \begin{center} \end{center}
\includegraphics[width=0.44 \textwidth,origin=c,angle=0]{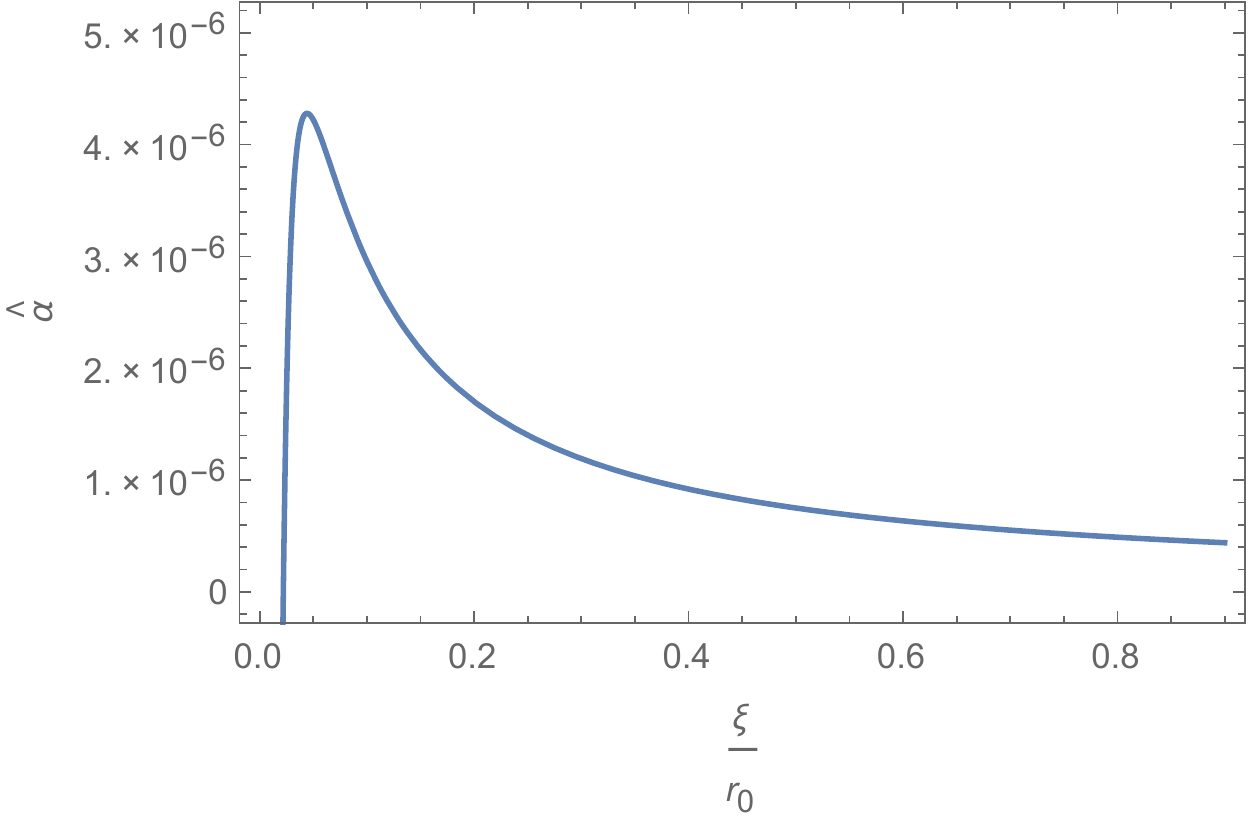}
\caption{\label{fig:p3} $f(R)$ deflection angle profile obtained for ${\delta\approx O(10^{-6})}$ with the $r_0=10^{2.650}$ kpc  for the galaxy NGC 5533. The deflection angle increases with mass concentration and decreases in the dynamical $f(R)$ background geometry. The value of closest approach is approximately 22 kpc.}\label{f3}
\end{figure}
\begin{figure}[!h]
\centering  \begin{center} \end{center}
\includegraphics[width=0.44 \textwidth,origin=c,angle=0]{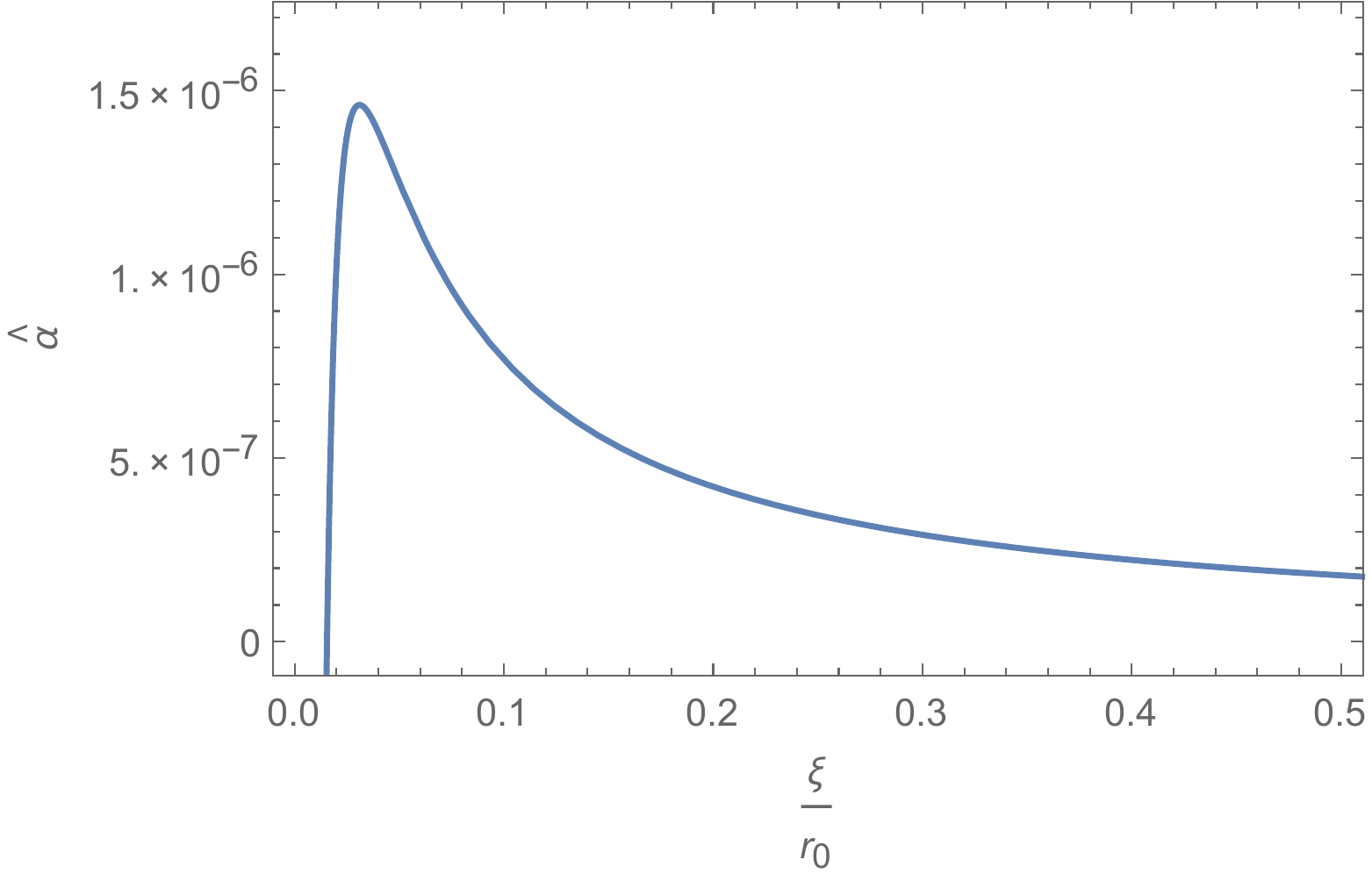}
\caption{\label{fig:p4} $f(R)$ deflection angle profile obtained for ${\delta\approx10^{-6}}$ with the $r_0=10^{2.215}$ kpc for the galaxy NGC 4138. The deflection angle increases with mass concentration and decreases in the dynamical $f(R)$ background geometry. The value of closest approach is approximately 6 kpc. }\label{f4}
\end{figure}
Fig. \textcolor[rgb]{0.00,0.00,1.00}{\ref{f3}} and Fig. \textcolor[rgb]{0.00,0.00,1.00}{\ref{f4}} represent the behaviour of light deflection angle in $f(R)$ background. It decreases in the $f(R)$ background geometry or scalaron halo region. The values of closest approach resembles approximately with the optical radius of the respective galaxies measured out to the isophotal level.
For the insightful interpretation of the magnitudes of $f(R)$ lensing angle plots, we compare them with the lensing angle profiles obtained according to the generalized pseudo-isothermal model in the next section.

\section{\label{5} Generalized pseudo-isothermal dark matter model and lensing profile}

This galaxy model is proposed according to the demand of observations of the HI profile in the outer region of typical galaxy. Fitting of the rotational velocity profiles in that region is usually done by assuming the spherical dark matter halo concentric with the baryonic matter of the galaxy. Such model has the following density profile of matter distribution as a function of distance
explored in a generalized way in \textcolor[rgb]{0.00,0.00,1.00}{\cite{b34}} as
\begin{eqnarray}
\rho(r)=\rho_{0}e^{-(\frac{r}{r_c})^2}\label{a10},\end{eqnarray}
where $\rho_{0}$ is the halo core density and the core radius is $r_c$, for the isolated system whose metric element is
\begin{eqnarray}
ds^{2}= -e^{a(r)}dt^{2} + e^{b(r)}dr^2 + r^2 d\theta^{2} + r^2 \sin^{2}\theta d\phi^{2} \label{a11}.\end{eqnarray}

\begin{figure}
\centering  \begin{center} \end{center}
\includegraphics[width=0.440 \textwidth,origin=c,angle=0]{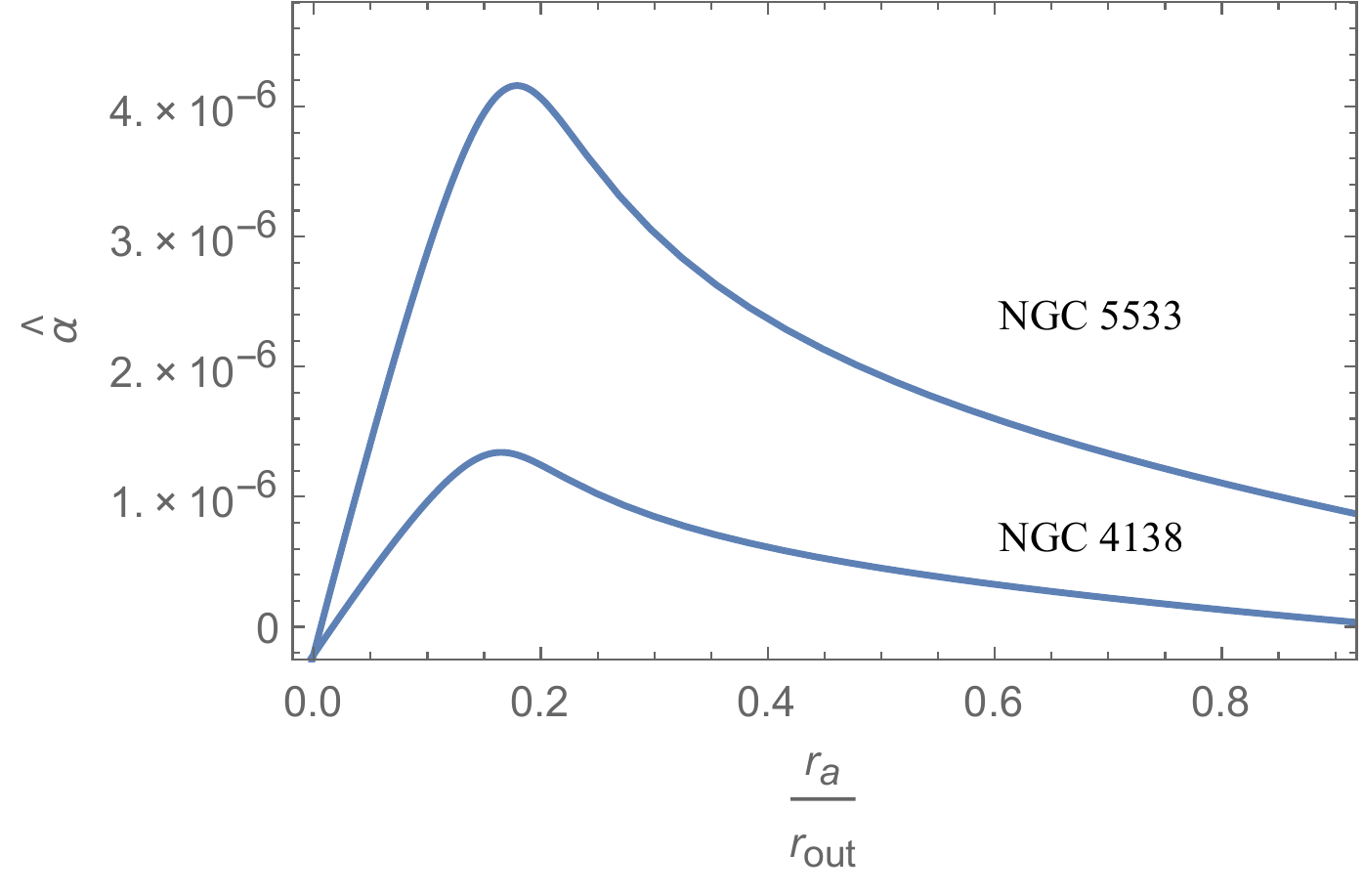}
\caption{\label{fig:p5} Light deflection angle curve for NGC 5533 and NGC 4138  plotted with generalised pseudo-isothermal dark matter density profile.  The   deflection angle increases up to the core radius and  decreases in the halo of dark matter background  \textcolor[rgb]{0.00,0.00,1.00}{\cite{b34}}.}\label{f5}
\end{figure}

Since, we work in dynamical modified $f(R)$ cosmological background, therefore to model the galaxy halo for the flatness behaviour of the rotational velocity profiles beyond the optical disk, we rely on their lensing angle profile in $f(R)$ background as because it independently probes the mass concentration (baryonic and non-baryonic) without relying on its dynamical state or nature. We then compare $f(R)$ lensing profile with the generalised pseudo-isothermal dark matter model for the galaxy.
The usual lensing angle profile according to equation \textcolor[rgb]{0.00,0.00,1.00}{(\ref{a10})} in the weak field limit is \textcolor[rgb]{0.00,0.00,1.00}{\cite{b34,b37}}
\begin{eqnarray}
\hat{\alpha}=16\pi G_N\rho_0\left(\frac{r_{out}}{r_a}\right)^2\times \nonumber\\ \left[-\frac{1}{2} e^{-(\frac{r_a}{r_{out}})^2}+\frac{\sqrt{\pi}}{4}\left(\frac{r_{out}}{r_a}\right)Erf\left(\frac{r_a}{r_{out}}\right)\right]\label{a12},
\end{eqnarray}
where ${r_a}$ is the radius of the closest approach to the centre of the galaxy, ${r_{out}}$ is the outer radius which is the last observed data point for the study of rotation curve or upto which the rotation curve is traced approximately flat and  $\rho_0$= 10$^{-14}$ kg m$^{-3}$ \textcolor[rgb]{0.00,0.00,1.00}{\cite{b34,b37,b38}}.
With the above specifications,  the plot of light deflection angle profile for the density profile given by equation \textcolor[rgb]{0.00,0.00,1.00}{(\ref{a10})} w.r.t $\frac{r_a}{r_{out}}$ according to equation \textcolor[rgb]{0.00,0.00,1.00}{(\ref{a12})} is shown in Fig. \textcolor[rgb]{0.00,0.00,1.00}{\ref{f5}} for NGC 5533 and NGC 4138.\\

Thus, on comparing the lensing magnitudes of plots (Figs. \textcolor[rgb]{0.00,0.00,1.00}{\ref{f3}} and \textcolor[rgb]{0.00,0.00,1.00}{\ref{f4}} with Fig. \textcolor[rgb]{0.00,0.00,1.00}{\ref{f5}}) for the said galaxies, we interpret that the magnitudes are in approximate agreement but with a shift in the peak value occurs (in case of Figs. \textcolor[rgb]{0.00,0.00,1.00}{\ref{f3}} and \textcolor[rgb]{0.00,0.00,1.00}{\ref{f4}}) because we consider the halo of scalaron cloud and do not consider any dark matter halo profile as compared to the position of peak   in Fig. \textcolor[rgb]{0.00,0.00,1.00}{\ref{f5}}.  It is possible that such scalaron density  against the background of   the high energy density regions in   the galactic environment  becomes   high because of the chameleon mechanism and thus the Compton wavelength of the scalaron profile becomes much smaller.  Hence, such features of the scalaron might explain the anomaly observed in the profile of lensing angle plots. Therefore, the baryonic distribution among the two galaxies decreases beyond their closest approach in $f(R)$ background without any dark matter. Hence, we have a declining nature of deflection angle in the halo region with the distance which is clear from the Fig. \textcolor[rgb]{0.00,0.00,1.00}{\ref{f3}} and Fig. \textcolor[rgb]{0.00,0.00,1.00}{\ref{f4}}.

\section{\label{6}Summary and Discussions}
The gravitational lensing profile is generally used for the mass measurement of astrophysical objects and in this work,  we utilized  it to test the predictions of modified gravity theory.   Thus,  we introduce  a formalism to study the effects of clustered dark matter problem in $f(R)=\frac{R^{1+\delta}}{R_c^{\delta}}$ type model at the galactic scales via obtaining the dependent lensing angle profiles of typical nearby galaxies on the $f(R)$ rotational velocity profiles. We use  the common point like effective potential in $f(R)$ background, motivated by  the work of  Capozziello et al.,  \textcolor[rgb]{0.00,0.00,1.00}{\cite{b24,b32}}, in order to study the light deflection angle and explore it for two nearby controversial galaxies in $f(R)$ background. The magnitude of light deflection is obtained via exploring the $f(R)$ characteristic length scale ( which  appeared for power-law $f(R)$ modification due to Noether's  symmetry) from the effective $f(R)$ rotational velocity for the two galaxies with $\delta\approx 10^{-6}$ at galactic scales \textcolor[rgb]{0.00,0.00,1.00}{\cite{b31}}. The profile of effective $f(R)$ rotational velocity for the galaxies is discussed in Fig. \textcolor[rgb]{0.00,0.00,1.00}{\ref{f1}} and Fig. \textcolor[rgb]{0.00,0.00,1.00}{\ref{f2}} whereas in Fig. \textcolor[rgb]{0.00,0.00,1.00}{\ref{f3}} and Fig. \textcolor[rgb]{0.00,0.00,1.00}{\ref{f4}}, the profile of  net light deflection angle is explored. We compare our obtained results of the net $f(R)$ light deflection angle with the generalized pseudo-isothermal dark matter model for a  galaxy and is interpreted to provide  a nice agreement. It may also be interpreted from the plots of light deflection angle in $f(R)$ background that the said galaxies have such closest approach whose value is approximately matched with the isophotal optical radius and beyond that the baryonic distribution decreases in $f(R)$ background and explain the results without any dark matter. Hence, we have a declining profile of deflection angle beyond the closest approach in the scalaron cloud region (or $f(R)$ background region). Thus, we can regard such scalaron cloud region to mimic as dark matter halo.

  Our result of $f(R)$ light deflection angle with $\delta\approx O(10^{-6})$  is in  close agreement  with the result obtained with the generalised pseudo-isothermal  model. The magnitude of effective $f(R)$ lensing angle is  $\hat{\alpha}_{Net}\approx 4.2\times 10^{-6}$ and $\hat{\alpha}_{Net}\approx 1.5\times 10^{-6}$ for NGC 5533 and NGC 4138,  respectively,   with the corresponding   approximate closest approach at  about  22 kpc and 6 kpc in $f(R)$ background. Thus, we diagnose our $f(R)$ model parameter  $\delta$  with   lensing angle profile at galactic scales. Also, the shift in the peaks  may be attributed to the scalaron  over-densities  which  rises  in the high energy density regions  of galactic surroundings   because of the chameleon mechanism and thus the Compton wavelength of the scalaron profile becomes much smaller. Such interesting features may possibly  explain the offset of peak profiles observed in the  Bullet cluster (1E0657-56), Abell 520 system etc. Furthermore, as an important implication of our analysis of galactic dynamics in $f(R)$ background, it may be possible to think about the extent of such scalaron halo cloud to mimic as dark matter halo in $f(R)$ background according to \textcolor[rgb]{0.00,0.00,1.00}{\citep{b031}}. Katsuragawa et al., have investigated  the  particle picture of scalaron as a dark matter candidate.  The emergence of $f(R)$ characteristic length scale may predict the extent of scalaron halo surrounding the galaxy since the trace of light deflection angles follows the same $f(R)$ characteristic lengths as obtained for velocity curves and the deflection magnitudes agree   closely with  the  generalised pseudo-isothermal model. For the two galaxies (NGC 5533 and NGC 4138), we have the value of such approximate $f(R)$ characteristic length as $446.600$ kpc and $ 177.828$ kpc which might represent their halo size.  As the halo size also evolves which indicates the unstable nature of dark matter halo, so the $f(R)$ gravity theory can be used for such study and we plan to study such implications in our future work.

Thus, from our analysis, it is reasonable to expect that we can explain some of the galactic dynamics effects probably attributed to the presence of large amount of dark matter, although further work must be carried out to improve on the point source for galaxies, before strict conclusion can be drawn.\\
Hence, with the single effective point like potential and with $\delta\approx O(10^{-6})$, we have explored the combined observations of galaxies i.e., their light deflection angle through the rotational velocity profiles in the $f(R)$ background.
We hope that, such combined study might shed new light on the distribution of clustered dark matter halo (or scalaron cloud in $f(R)$ background) surrounding the galaxy.
\section*{Acknowledgments}
Authors thank  IUCAA, Pune, for the computational  facilities extended  where   a part of the present work was completed under the associateship programme. VKS also  thanks     Varun Sahni for hosting the visit to IUCAA, Pune  and several  useful discussions.


\end{document}